\documentclass[useAMS,usenatbib]{mn2e}
\usepackage[dvips]{graphicx}
\usepackage{epstopdf}
\usepackage{times}
\usepackage{amsmath}
\usepackage{color}
\usepackage{ulem}
\def\simgt{\lower.5ex\hbox{$\; \buildrel > \over \sim \;$}}
\def\simlt{\lower.5ex\hbox{$\; \buildrel < \over \sim \;$}}
\newcommand{\aap}{A\&A}

\newcommand{\mnras}{MNRAS}

\newcommand{\apj}{ApJ}
\newcommand{\apjl}{ApJ}

\newcommand{\aj}{AJ}

\title[Binary star evolution]{Evolution of Binary Stars in Multiple-Population Globular Clusters - II. Compact Binaries}
\author[J. Hong et al.]  {Jongsuk Hong$^1$\thanks{E-mail: hongjong@indiana.edu}, Enrico Vesperini$^1$, Antonio Sollima$^2$, Stephen L. W. McMillan$^3$,\\ \newauthor
Franca D'Antona$^4$ and Annibale D'Ercole$^2$\\
  $^1$Department of Astronomy, Indiana University, Bloomington, IN, 47401, USA\\
  $^2$INAF - Osservatorio Astronomico di Bologna, via Ranzani 1, I-40127 Bologna, Italy\\
  $^3$Department of Physics, Drexel University, Philadelphia, PA 19104, USA\\
  $^4$INAF - Osservatorio Astronomico di Roma, via di Frascati 33, I-00040 Monteporzio, Italy\\
}

\begin{document}
\newcommand{\fbfg}{f_{\rm b,FG}}
\newcommand{\fbsg}{f_{\rm b,SG}}
\newcommand{\xg}{x_{\rm g,0} }
\newcommand{\trel}{\tau_{\rm rh,0} }
\newcommand{\rglob}{R_{\rm glob}}
\date{Accepted ... Received ...; in original form ...}
\maketitle

\label{firstpage}

\begin{abstract}
We present the results of a survey of $N$-body simulations aimed at exploring the evolution of compact binaries in multiple-population globular clusters. 

We show that as a consequence of the initial differences in the structural properties of the  first-generation (FG) and the second-generation (SG) populations and the effects of dynamical processes on binary stars, the SG binary fraction decreases more rapidly than that of the FG population. The difference between the FG and SG binary fraction is qualitatively similar to but quantitatively smaller than that found for wider binaries in our previous investigations.

The evolution of the radial variation of the binary fraction is driven by the interplay between  binary segregation, ionization and ejection. Ionization and ejection counteract in part the effects of mass segregation but for compact binaries the effects of segregation dominate and the inner binary fraction increases during the cluster evolution. We explore the variation of the difference between the FG and the SG binary fraction  with the distance from the cluster centre and its dependence on the binary binding energy and cluster structural parameters.

The difference between the binary fraction in the FG and the SG populations found in our simulations is consistent with the results of observational studies finding a smaller binary fraction in the SG population.

\end{abstract}

\begin{keywords}
globular clusters:general, stars:chemically peculiar.
\end{keywords}

\section{Introduction}
\label{sec:intro}
A number of observational studies
have shown that globular clusters
host multiple stellar populations. Different stellar populations are
characterized by different abundances in light elements such as Na, O,
Mg, Al and in some cases also by significant differences in helium
(see e.g. Carretta et al. 2009a, 2009b, Gratton et al. 2012, Piotto et
al. 2015 and references therein)

As a consequence of this discovery new questions arose and many fundamental issues concerning the formation and dynamical history of globular clusters require new investigations.
Most questions concerning multiple-population clusters are still unanswered and significant observational and theoretical efforts are needed to make progress towards a complete understanding of these stellar systems.

One of the key questions concerns the source of processed gas out of
which second-generation stars (hereafter SG) formed;
although no consensus has been reached on this fundamental issue all models
proposed so far agree that SG stars should form in the central regions
of a more diffuse first-generation (hereafter FG) system (see
e.g. D'Ercole et al. 2008, Decressin et al. 2007, Bastian et al. 2013\footnote{
Notice that in the context of the study of Bastian et al. 2013 there
are no different generations of stars and the term 'populations' is more
appropriate to describe stars with different chemical properties;
hereafter, for convenience, we will continue to refer to SG and FG to
indicate stars with different chemical properties.})
and several observational studies have found clusters in which the
initial central concentration of the SG population has not been erased
by the subsequent dynamical evolution (see e.g. Sollima et al. 2007;
Bellini et al. 2009; Lardo et al. 2011; Nataf et al. 2011; Milone et al. 2012; Beccari et
al. 2013; Johnson \& Pilachowski 2012; Cordero et al. 2014; Kucinskas,
Dobrovolskas \& Bonifacio 2014). Differences in the FG and SG
velocity anisotropy which might arise as a consequence of the initial spatial differences have also been observed in the first proper-motion studies of multiple-population clusters (Richer et al. 2013, Bellini et al. 2015).

Binary stars play an important role in the dynamical evolution of
globular clusters and understanding their evolution in
multiple-population clusters is one of the important issues to
address in the study of the dynamical evolution of these systems. 
 Specifically, here we explore the implications of the differences in the SG and FG
spatial distributions for the evolution and survival of their binary
populations.

An
initial study of the differences in the evolution of the FG and SG
binary populations  based on a combination of $N$-body simulations and
analytical calculations of binary ionization was carried out by
Vesperini et al. (2011). More recently, in  Hong et al. (2015;
hereafter paper I) we have carried out an extensive survey of $N$-body
simulations to explore the evolution of moderately hard binaries in
multiple-population clusters. Both these investigations concluded that
differences in the initial structural properties of the FG and SG
populations may leave a significant fingerprint in the binary fraction
and binding energy distribution  of different stellar populations. 

Specifically, these studies showed that SG binaries, as a consequence
of their initial central 
concentration,  are more significantly affected by dynamical processes
and predicted 
that the fraction of SG binaries (in the moderately hard regime)
should be smaller than that of the FG population. On the observational
side, D'Orazi et al. (2010) and Lucatello et al. (2015) found trends in the FG and SG binary fractions
consistent with the results of these theoretical investigations.  

 In this paper we extend the investigation of paper I to the regime of
compact binaries. For more compact binaries the role played by different dynamical processes and the
outcomes of binary interactions are different from those of the
moderately hard binaries considered in paper I and it is therefore
necessary to extend the survey of simulations presented in paper I.
Specifically, for the compact binaries considered in this paper
binary-single interactions do not contribute to the process of binary
ionization which is  driven mainly by  binary-binary interactions;
both binary-binary and binary-single 
interactions may result in large recoil speeds for the interacting
stars and  in their ejection from the cluster. Finally since the
interaction rate of compact binaries is non-negligible only in the
cluster innermost regions, the evolution of binary properties is
driven also by the timescale on which internal two-body
relaxation drives binary segregation towards the cluster centre. Here
 we explore the implications of the differences in the initial spatial structure of the
FG and SG population, and in the rate of dynamical processes driving the
evolution of FG and SG binary properties and spatial segregation; our analysis is aimed at determining whether these difference leave any fingerprint in the
current properties of compact binaries.

The outline of this paper is the following. In section 2, we describe the method and the
initial conditions adopted in our simulations. In section 3, we present
the results concerning the evolution of the global binary fraction
(section 3.1), the evolution of the binding energy distribution
(section 3.2), and the radial variation of the binary fraction
(section 3.3). Our conclusions are summarized in section 4.

\section{Methods and Initial Conditions}
\begin{table}
  \begin{center}
  \caption{Initial parameters for all models.}
  \begin{tabular}{l c c c c c c}
  \\
    \hline
    \hline
    Model id. & $N$ & $\frac{R_{\rm h,FG}}{R_{\rm h,SG}}$ & $f_{\rm b,0}$ & $x_{\rm g,0}$ \\
    \hline
MPr5f03x50     & 20,000 & 5 & 0.03 & 50\\
MPr5f03x100     & 20,000 & 5 & 0.03 & 100\\
MPr5f03x400     & 20,000 & 5 & 0.03 & 400\\
MPr5f03x800     & 20,000 & 5 & 0.03 & 800\\
MPr10f03x50     & 20,000 & 10 & 0.03 & 50\\
MPr10f03x100     & 20,000 & 10 & 0.03 & 100\\
MPr10f03x400     & 20,000 & 10 & 0.03 & 400\\
MPr10f03x800     & 20,000 & 10 & 0.03 & 800\\
MPr5f1x50     & 20,000 & 5 & 0.1 & 50\\
MPr5f1x100     & 20,000 & 5 & 0.1 & 100\\
MPr5f1x400     & 20,000 & 5 & 0.1 & 400\\
MPr5f1x800     & 20,000 & 5 & 0.1 & 800\\
MPr5f1x3-800     & 20,000 & 5 & 0.1 & 3-800$^a$\\
MPr10f1x50     & 20,000 & 10 & 0.1 & 50\\
MPr10f1x100     & 20,000 & 10 & 0.1 & 100\\
MPr10f1x400     & 20,000 & 10 & 0.1 & 400\\
MPr10f1x800     & 20,000 & 10 & 0.1 & 800\\
    \hline
SPf1x50     & 20,000 & - & 0.1 & 50\\
SPf1x100     & 20,000 & - & 0.1 & 100\\
SPf1x400     & 20,000 & - & 0.1 & 400\\
SPf1x800     & 20,000 & - & 0.1 & 800\\
    \hline
    \label{tbl1}
  \end{tabular}
  \end{center}
\begin{flushleft}
$N=N_s+N_b$ is the total number of single,$N_s$, and binary, $N_b$, particles.\\
$f_{\rm b,0}=N_b/(N_s+N_b)$ is the initial binary fraction.\\
$\frac{R_{\rm h,FG}}{R_{\rm h,SG}}$ is the ratio of the FG to the SG initial  half-mass radii.\\
All the SP models refer to single-population systems.\\
$\xg$ initial hardness parameter (see section 2 for definition).\\
$^a$ A uniform distribution in binding energy is assumed.\\
\end{flushleft}
\end{table}

We performed a survey of $N$-body simulations using the GPU-accelerated version of {\sc nbody}6 code (Nitadori \& Aarseth 2012) on the {\sc big red ii} GPU cluster at Indiana University. 

In the simulations of multiple-population (MP) systems, the initial FG and SG subsystems are both modeled as King models (King 1966) with central dimensionless potential equal to $W_0 = 7$, each  with the same total mass but different spatial scales. Following the results of multiple-population cluster formation models (see e.g. D'Ercole et al. 2008) we start with a SG subsystem initially more centrally concentrated in the inner regions of a more diffuse FG system.
This study and the simulations discussed in this paper are focused on the long-term evolution of clusters driven by two-body relaxation and starting after the early phases of evolution during which the cluster might have lost a large fraction of the FG population (see D'Ercole et al. 2008).
We have explored systems with values of the ratio of the half-mass radii of the FG to the SG system, $R_{\rm h,FG}/R_{\rm h,SG}$, equal to 5 and 10. 
We have also explored the evolution of single-population (SP) systems as a reference system modeled as single King model with $W_0=7$ and a total mass, tidal radius, and number of binaries equal to those of the MP systems with 10 per cent binary fraction.

All the systems are initially tidally truncated and the effects of the tidal field of the host galaxy, modeled as a point mass, are included. Particles moving beyond a radius equal to two times the tidal radius are removed from the simulation.

For all our models, the total number of binary and single stars $N \equiv N_{\rm b} + N_{\rm s}=20,000$ where $N_{\rm b}$ and $N_{\rm s}$ are the number of binaries and single stars respectively. All stars in our models have the same mass.
The total number of particles depends on the initial binary fraction, $f_{\rm b,0}=N_{\rm b}/(N_{\rm s}+N_{\rm b})$, and is equal to $N_{\rm tot} \equiv 2N_{\rm b} + N_{\rm s}$.
We have explored systems with binary fractions, $f_{\rm b,0} = $0.03 and 0.1. Thus, the total number of particles are 20,600 and 21,000 for $f_{\rm b,0} = $0.03 and 0.1, respectively.

To characterize the initial hardness of primordial binaries in our simulations we use the parameter, 
$\xg \equiv E_b/(m \sigma_{\rm SP}^2)$ where $E_b$ is the absolute value of the binary binding energy, and $\sigma_{\rm SP}$ is the 1-D velocity dispersion of all stars in the SP system.
The range of values of  $\xg$ and the fraction of binaries considered in this investigation are summarized in Table 1. 
In order to characterize the extent of stochastic variations in our results we have repeated the simulations for the MPr5f03x800 and the MPr10f03x800 systems with six different random realizations of the initial conditions.

 The hardness
parameter allows us to characterize the possible role and
evolution of a binary during the dynamical history of a cluster (see e.g. Heggie \& Hut 2003); the
actual physical properties of a binary corresponding to a given value
of the hardness parameter depend on
the masses of the binary components, the binary semi-major axis (or
period) and the velocity dispersion of the cluster.
Assuming, for example, a mass for single stars equal to 0.5
$m_{\odot}$, a value for the velocity dispersion $\sigma$
equal to 5 km/s, a binary with equal-mass components each with a mass equal
to 0.8 $m_{\odot}$, the range of values of $x$ explored in paper I and
in this paper  ($3<x<800$) would correspond to periods (semi-majors
axes) ranging from about $16.5$ years (about $7.5$ AU) to $4\times 10^{-3}$ years (about $3 \times 10^{-2}$ AU). 
We emphasize that these are just some indicative values but they will be different for clusters and/or binaries with different properties.

\section{Results}

\subsection{Evolution of the global binary fraction}

\begin{figure}
  \includegraphics[width=84mm]{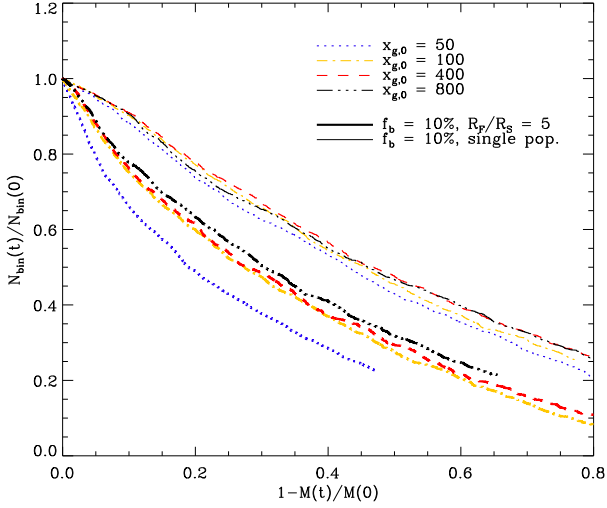}
  \caption{Evolution of the total number of binaries (normalized to
    the initial number of binaries $N_{\rm bin}(0)$)  as a function of
    the mass lost by the cluster for the MP5r5 and the SP simulations and
    for different initial values of the binary hardness parameter. }
  \label{fig_tnumbm}
\end{figure}

Fig. \ref{fig_tnumbm} shows the evolution of the total number of binaries for the MP models MPr5f1x50, MPr5f1x100, MPr5f1x400, MPr5f1x800 and the corresponding SP models SPf1x50, SPf1x100, SPf1x400, SPf1x800. 
For compact binaries like those considered in this study, both relaxation-driven evaporation and  disruption and ejection  due to  binary-single and binary-binary encounters in the cluster central regions are important in driving the evolution of the number of binaries.

In order to separate the effects due to evaporation from those related to binary interactions and to better illustrate the implications of the differences in the structural properties of MP and SP, clusters we have plotted the evolution of the binary number as a function of the total fraction of the initial cluster mass remaining at any given time. By comparing the fraction of binary stars in MP and SP systems at times when they have lost the same fraction of their initial mass we can better highlight how the evolution of the number of binaries is affected by differences in the rate of dynamical processes (ionization and ejection) related to the different structural properties of the MP and SP systems.  The presence of a compact SG subsystem in the MP system structure leads to an enhancement of binary interactions which, in turn, lead to an enhancement in the binary disruption in MP systems. Indeed, as shown in Fig. \ref{fig_tnumbm}, the fraction of binaries decreases more rapidly in the MP simulations than in the SP ones.

\begin{figure}
  \includegraphics[width=84mm]{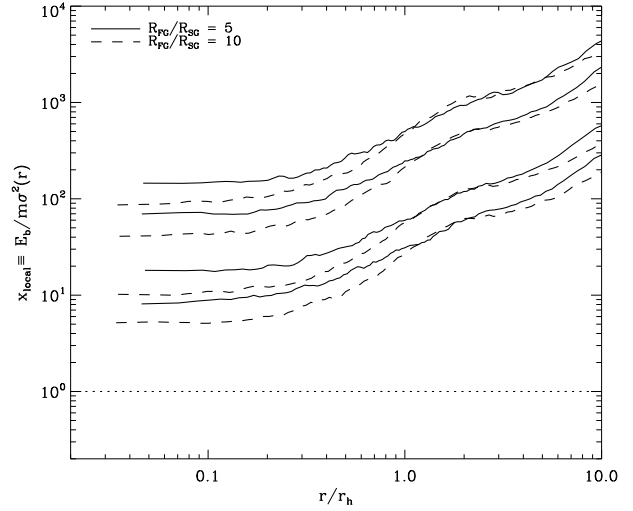}
  \caption{Radial profiles of the initial local hardness $x_{\rm local}(r)$ for simulations MPr5f1 (solid lines) and MPr10f1 (dashed lines).
  Radius is normalized to the half-mass radius. Different profiles
  correspond to values of  $\xg$
  equal to (from the bottom to the top line) 50, 100, 400, and 800.}
  \label{fig_hard}
\end{figure}

Fig. \ref{fig_hard} shows  the radial profile of the local hardness of primordial binaries defined as the ratio of binary binding energy to $m\sigma^2(r)$ where $\sigma(r)$ is the 1-D velocity dispersion at the radius $r$ for the MP models.
As shown in this figure, even in the central regions, the local hardness of the binaries considered in this study is significantly larger than one. This implies that binary stars are unlikely to be ionized by binary-single encounters. 
\begin{figure}
  \includegraphics[width=84mm]{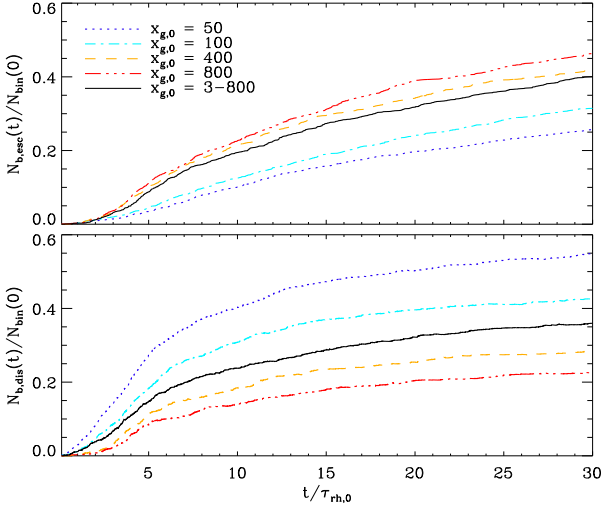}
  \caption{Time evolution of the number of escaped (top panel) and
    disrupted binaries (bottom panel) (normalized to the initial
    number of binaries $N_{\rm bin}(0)$) for the MPr5f1 simulations
    with different values of $\xg$. Time is normalized to the initial
    half-mass relaxation time $\tau_{\rm rh,0}$.}
  \label{fig_dis}
\end{figure}

In Fig. \ref{fig_dis} we show the time evolution of the number  of escaped or disrupted binaries for all the MPr5f1 simulations.
As shown in this figure, simulations with harder binaries are characterized by a higher escape rate and a lower disruption rate than those with less hard binaries. As expected, the fraction of binaries disrupted decreases as the hardness parameter increases.
Hard binaries are very rarely disrupted by binary-single interactions even in the centre where, as shown in Fig. \ref{fig_hard}, their hardness parameter is significantly larger than one.  Binary-binary interactions are the only interactions that can lead to binary disruption while both binary-single and binary-binary interactions can result in the ejection of the interacting binaries.
While binary ejection plays a minor role for the moderately hard binaries explored in paper I, it becomes increasingly important for harder binaries as the recoil speed after an interaction increases with the binding energy of the interacting binary. This is illustrated by the increase in the fraction of escaped binaries with the hardness parameter in the top panel of Fig. \ref{fig_dis}; in particular for the model MPr5f1x400, MPr5f1x800, and MPr5f1x3-800 the fraction of escaped binaries exceeds the fraction of disrupted binaries.

\begin{figure}
  \includegraphics[width=84mm]{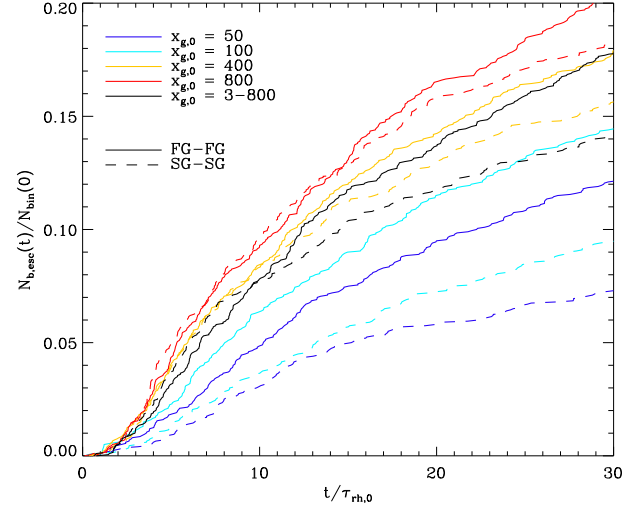}
  \caption{Time evolution of the number of FG (solid lines) and SG (dashed lines) escaped binaries (normalized to the initial total number of binaries $N_{\rm bin}(0)$) for MPr5f1 simulations with different initial $\xg$.}
  \label{fig_escp}
\end{figure}

Additional details on escaped binaries are presented in Fig. \ref{fig_escp}. 
In this figure, we show the time evolution of the number of escaped FG and SG binaries.
As discussed above,  binary escape is driven by evaporation and ejection;  binary evaporation is more important for FG binaries which, initially, are located preferentially in the cluster outer regions. On the other hand,  ejection is the result of binary interactions which are more frequent in the cluster inner regions and is therefore affecting predominantly SG binaries.

While the rate of binary escape is always larger for FG binaries, the ejection rate of SG binaries increases for increasing binary hardness.  The increase in the SG ejection rate leads to a decrease in the differences between the escape rate of SG and FG binaries for the largest values of $\xg$ considered.

\begin{figure*}
  \includegraphics[width=161mm]{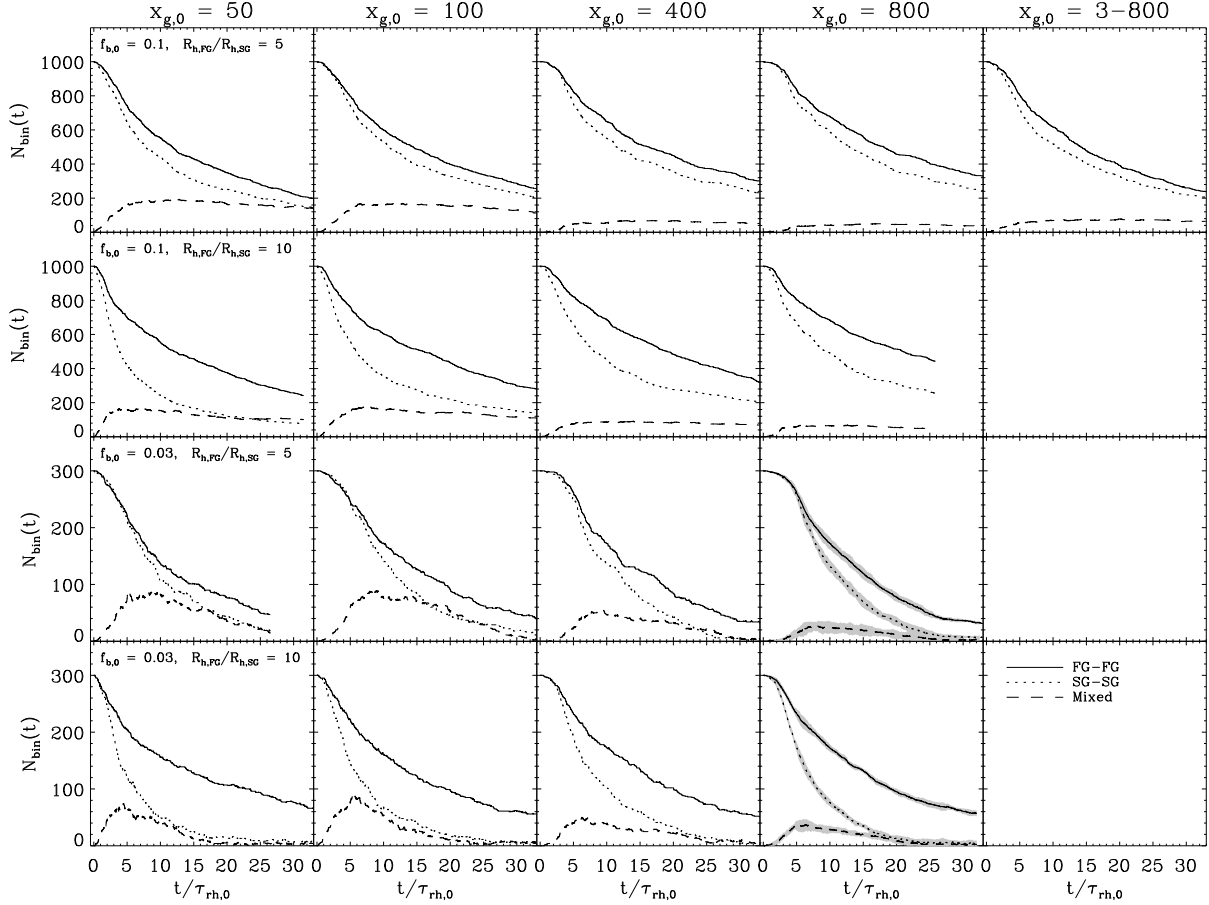}
  \caption{Time evolution of the number of FG (solid lines), SG (dotted lines) and mixed (dashed lines) binaries for 
  MPr5f1, MPr10f1, MPr5f03 and MPr10f03 from top to bottom respectively. Time is normalized to the initial half-mass relaxation time. 
  Each column shows the results of simulations with $\xg$ presented on the top of each column.
  For the simulations MPr5f03x800 and MPr10f03x800, the lines show the mean value of the number of binaries of six simulations with different random realizations of the initial conditions and the shaded area show the one sigma dispersion.}
  \label{fig_numb}
\end{figure*}

Fig. \ref{fig_numb} shows the time evolution of total number of SG, FG and mixed binaries  (hereafter we will refer to binaries with one FG component and one SG component as mixed binaries) for all the MP models.
As expected for given values of $R_{\rm h,FG}/R_{\rm h,SG}$ and $f_{\rm b,0}$ the number of binaries decreases more rapidly for less hard binaries.

For all the systems investigated we find that the number of SG
binaries declines more rapidly than that of FG binaries. 
Differences in the time
evolution of SG and FG binaries are more marked in the MPr10f1 models.
As a consequence of the higher density and velocity dispersion of the
MPr10f1 models, interactions leading to binary ionization and ejection
can start earlier and lead to a more rapid decrease in the binary number
than the MPr5f1 models. 

Mixed binaries are produced in binary-single or
binary-binary encounters resulting in the exchange of one of the
binary components with a star of a generation different from
that of the two original binary components. Mixed binaries can, in turn, exchange one
of their components and become again an SG or an FG binary.
When the number of mixed binaries becomes comparable to that
of SG and FG binaries (particularly in the central regions where
most exchange encounters are expected to occur) the rates of these different exchange events also become comparable and the evolution of the number of mixed binaries follow the decline of the number of SG and FG binaries.

\begin{figure}
  \includegraphics[width=84mm]{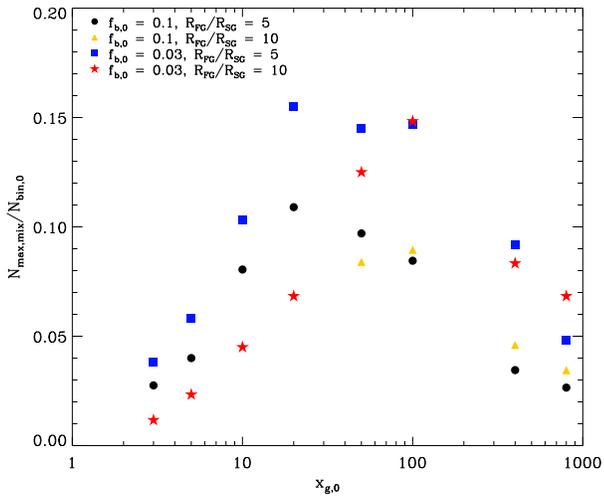}
  \caption{Maximum number of mixed binaries obtained from Fig. \ref{fig_numb} as a function of $\xg$. 
  Different symbols show the results of simulations MPr5f1, MPr10f1, MPr5f03 and MPr10f03. 
  The maximum number of mixed binaries for moderately-hard binaries ($3<\xg<20$) are from the figure 4 of paper I.}
  \label{fig_mix}
\end{figure}

Interesting information about the dynamics of binary interactions is contained in the formation of mixed binaries.
Fig. \ref{fig_mix} shows the maximum number of mixed binaries during the cluster evolution for all
the simulations discussed in this paper and those presented in paper I
(see Fig. 4 of paper I).
The number of mixed binaries is expected to decrease for softer
binaries because they can be easily ionized during the interactions
needed to produce a mixed binary. Similarly the number of mixed
binaries is expected to decrease for very hard binaries since the
large recoil speed acquired after the interactions producing a mixed
binary is likely to lead to the mixed binary ejection.
Fig. \ref{fig_mix} confirms this general expectation and shows that the maximum number of mixed binary is achieved for intermediate values of the binary hardness ($20<\xg<100$).

\subsection{Evolution of the binary binding energy}
\begin{figure}
  \includegraphics[width=84mm]{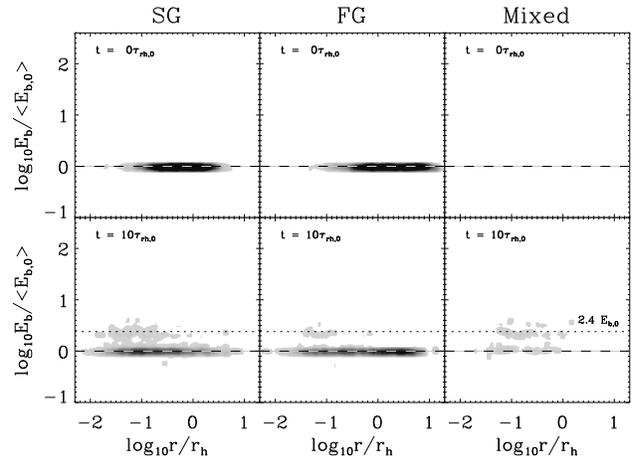}
  \caption{Distribution of SG (first column), FG (second column) and
    mixed (third column) binaries for the model MPr5f1x400 on the
    $E_{\rm bin}-r/r_{\rm h}$ plane. 
  The binding energy is normalized to the initial binding energy $E_{\rm bin,0}$.
The horizontal dashed line indicates the initial value of the binding
energy, $E_{\rm bin,0}$; the horizontal dotted line indicates the value of
binding energy corresponding to $2.4 E_{\rm bin,0}$ (see section 3.2 for discussion).
 Data from three snapshots around a given time in each panel are combined for this figure. }
  \label{fig_deb400}
\end{figure}

Fig. \ref{fig_deb400} shows the distribution of binaries on $r-E_{\rm bin}$ plane at $t/\trel=$ 0 and 10 for the simulation MPf1r5x400. 

The panels for $t/\trel=10$ clearly show two distinct groups of binaries with different binding energies. 
One of the two groups corresponds to binaries with binding energy equal to the initial value adopted in the simulation;
in order to investigate the origin of the other group,  we have tracked individual binary encounters and found that the group of binaries with a larger binding energy in Fig. \ref{fig_deb400} is the result of binary-binary encounters producing one harder binary and two single stars.
The binding energy of this group is in very good agreement with the predictions of the study of binary-binary encounters by Mikkola (1983).
Specifically, Mikkola (1983) found that in binary-binary encounters resulting in the disruption of one binary, the most likely outcome was one in which the surviving binary emerged from the encounter with its binding energy increased by an amount about equal to the sum of the binding energy of the disrupted binary plus 20\% of the total binding energy of the interacting binaries.
In the MPr5f1x400 simulation, all binaries have the same initial binding energy,$E_{\rm bin,0}$, and  this variation corresponds to a predicted final binding energy equal to $E_{\rm bin}\sim2.4E_{\rm bin,0}$  in very good agreement with what shown in Fig.\ref{fig_deb400}.

\subsection{Radial variation of the binary fraction}
\begin{figure}
  \includegraphics[width=84mm]{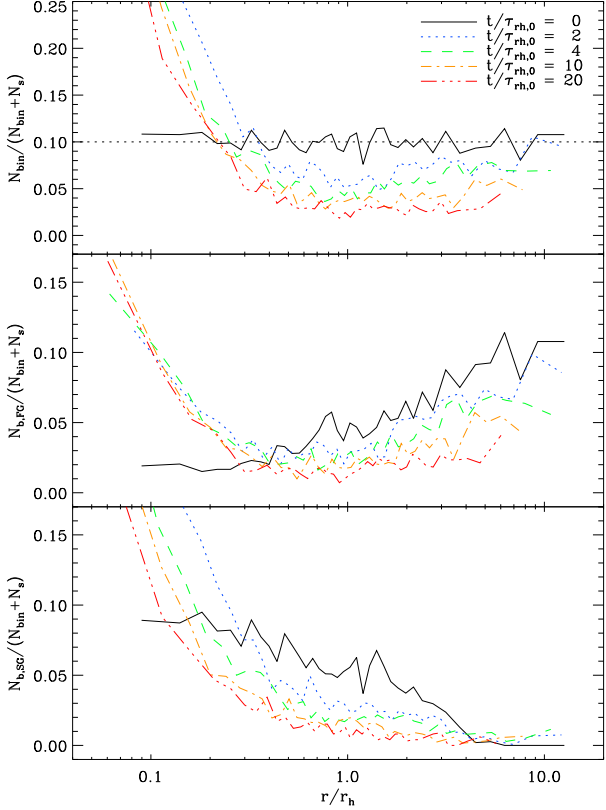}
  \caption{Time evolution of the radial profiles of the binary
    fraction (defined as the ratio of the number of binaries to the
    total number of binaries and single stars) for the model
    MPr5f1x3-800. Each profile is calculated  by combining three
    snapshots around the time indicated in the legend. Top, middle and
    bottom panels show, respectively,
    the profiles for all the binaries, for FG binaries,  and for SG
    binaries. The horizontal dashed line in the top
    panel shows the initial binary fraction. }
  \label{fig_bfr400}
\end{figure}

Fig. \ref{fig_bfr400} shows the radial profiles of the binary fraction
(defined as the ratio of the number of binaries for each population to
the total number of binaries and single stars in the cluster) for all
binaries, for FG binaries, and for SG binaries.  The low ionization
rate of the compact binaries we are considering facilitates the
central accumulation of binaries segregating from the outer regions
and results into an increasing binary fraction in the cluster inner
regions.  It is interesting to note that the evolved binary fraction radial profiles are in general qualitative agreement with those found
in the recent observational study of Milone et al. (2016).

\begin{figure}
  \includegraphics[width=84mm]{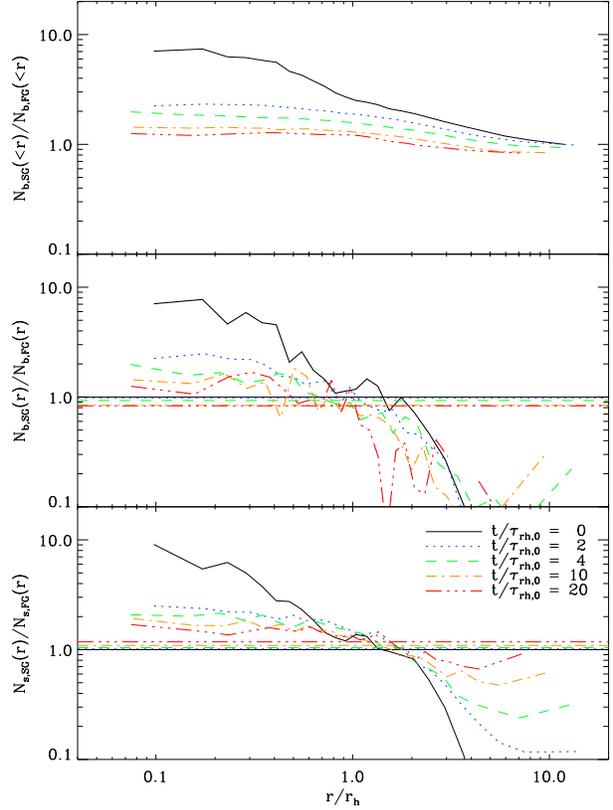}
  \caption{Time evolution of the radial profile of the cumulative number ratio of SG to FG binaries (top panel), the differential SG-to-FG binary number ratio (middle panel), 
and the differential number ratio of SG to FG single stars (bottom panel) for the model MPr5f1x3-800. 
Lines with different color and style correspond to the different times indicated in the legend, and the horizontal lines show the global SG-to-FG ratio for binaries (middle panel) and single stars (bottom panel).
(Each profile has been calculated using three combined snapshots
around the times indicated in the legend).}
  \label{fig_brr400}
\end{figure}

The three panels of Fig. \ref{fig_brr400} show the cumulative and the
differential radial profiles of SG-to-FG binary ratio and the
differential SG-to-FG ratio for single stars. 
The interplay among ionization, ejection and segregation towards the
inner regions results into a progressive decrease of the SG-to-FG
binary number ratio in the cluster inner regions.  
The global SG-to-FG binary number ratio decreases with time due to the
dominant effect of the preferential ionization or ejection of SG
binaries; it is interesting to note that this is opposite to what
happens to  the global SG-to-FG number ratio for single stars which
increases with time because for single stars the preferential evaporation of FG stars (see also Vesperini et
al. 2013) drives the evolution of this number ratio.

\begin{figure}
  \includegraphics[width=84mm]{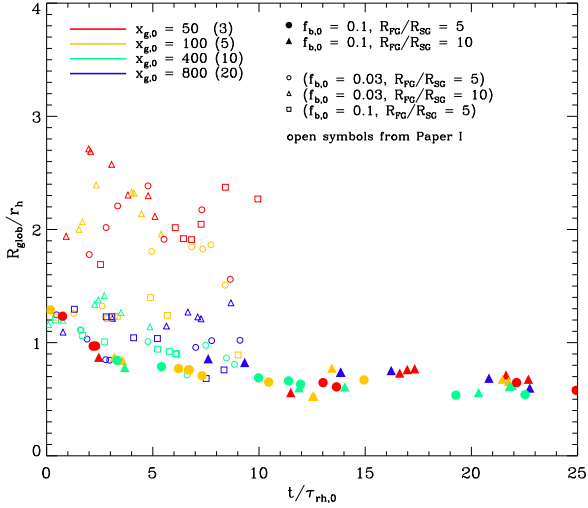}
  \caption{Time evolution of the radius, $\rglob$, where the local
    SG-to-FG binary ratio is equal to the global value of this ratio.
   Filled symbols show $\rglob$ in this study and open symbols are
   from paper I. Different colors correspond to different values of $\xg$ (numbers in parenthesis are for open symbols).}
  \label{fig_reql}
\end{figure}

As shown in Fig. \ref{fig_brr400}, the SG-to-FG binary number ratio depends on the distance from the cluster centre. It is important to identify the radius $\rglob$, where the local SG-to-FG ratio is equal to the global value of this ratio. 
In paper I, we have already shown that $\rglob$ depends on the initial hardness of binaries rather than the binary fraction or the cluster structure.
Fig. \ref{fig_reql} shows the time evolution of $\rglob$ for different simulations.
For moderately-hard binaries (e.g., $\xg = 3$ and 5) $\rglob \sim 2~r_{\rm h}$, and $\rglob$ decreases as $\xg$ increases.
 The simulations presented in this paper show that for compact binaries ($\xg \simgt 50$) $\rglob$ converges to 0.5$r_{\rm h}$.

\begin{figure}
  \includegraphics[width=84mm]{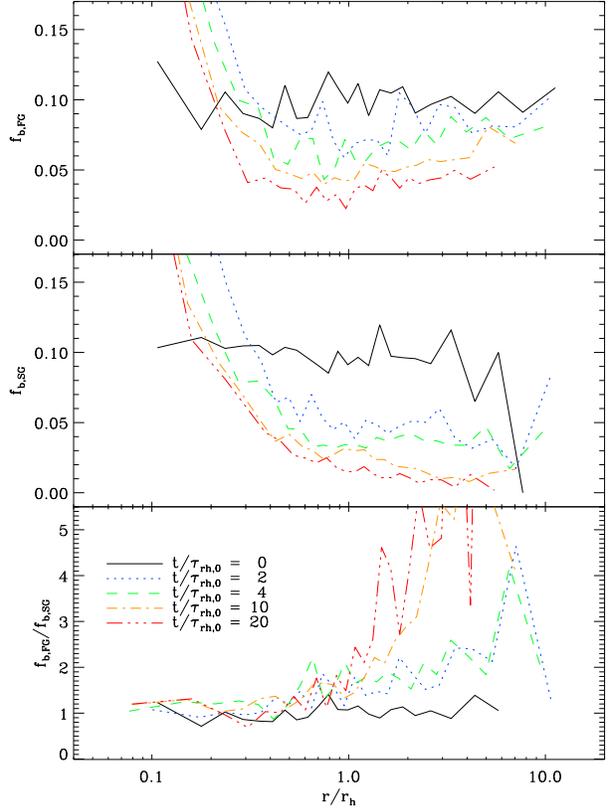}
  \caption{Time evolution of the radial profile of binary fraction
    $f_{\rm b}$ of each generation for the simulation MPr5f1x3-800. 
The  top panel shows the binary fraction for the FG (defined as ratio
of the number of FG binaries to the sum of the number of FG binaries
and FG single stars); middle panel shows the radial profile of the SG
population binary fraction.
  Bottom panel is the ratio of profiles of FG-to-SG binary fraction. 
Three snapshots are combined to calculate the radial profiles (except for $t/\trel=$10 and 20 for which 10 snapshots are used).}
  \label{fig_bfrr400}
\end{figure}

Finally in Fig. \ref{fig_bfrr400} we show the radial profiles of each population binary fraction 
defined as the ratio of the number of binaries of a given population to  the total number of single stars and binaries of the same population
(e.g., $f_{\rm b,FG} \equiv N_{\rm b,FG}/(N_{\rm s,FG}+N_{\rm b,FG})$, where $N_{\rm s,FG}$ and $N_{\rm b,FG}$ are the number of FG single stars and FG binaries, respectively); the radial profiles of the ratio of the binary fractions of each population,  $f_{\rm b,FG}/f_{\rm b,SG}$, are also shown in these figures.

The evolution of $\fbfg$ and $\fbsg$ in the innermost regions is determined by the interplay between the increase in the inner binary fraction driven by segregation and the binary fraction decrease  due to ionization and ejection.

In the innermost regions ($r<0.1-0.2 r_{\rm h}$) these dynamical processes
do not lead to significant differences in the incidence of binary
stars in each population. However outside the innermost regions the
incidence of binaries in the FG population  is always larger than that
of the SG population. The more extended initial spatial distribution
of the FG population implies that for the FG population there is a
larger outer repository of FG binary stars which slowly segregate
towards the cluster intermediate regions and slow down the decline of
the binary fraction due to the segregation of binaries in the
intermediate regions towards the cluster inner regions. The
combination of initial differences in the FG and SG spatial
distribution and the slow segregation of the outermost FG binary
population  results into a larger incidence of binaries in the FG
population in the intermediate regions.   
Moreover, as FG and SG single stars mix during the cluster dynamical evolution, 
the outer fraction of SG single stars increases and further contributes to the
decrease of the SG binary fraction in the cluster intermediate and outer regions.

\begin{figure}
  \includegraphics[width=84mm]{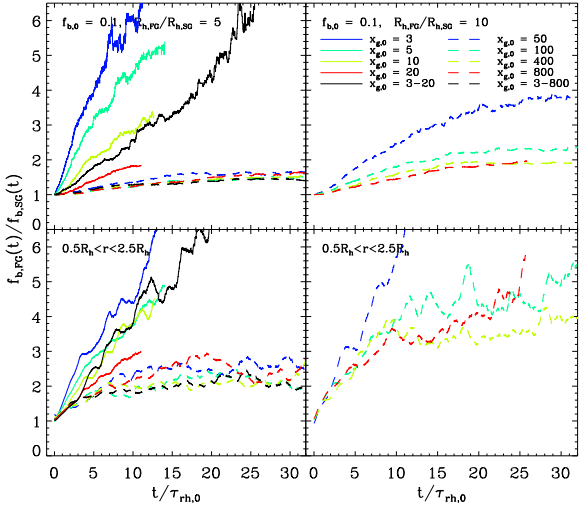}
  \caption{Time evolution of the global ratio of $f_{\rm b,FG}/f_{\rm
      b,SG}$ (top panels) and of $f_{\rm b,FG}/f_{\rm b,SG}$
    estimated at $0.5R_{\rm h}<r<2.5R_{\rm h}$ (bottom panels) for
    simulations MPr5f1 (first column) and MPr10f1 (second
    column). Dashed lines represent the results of this study and
    solid lines represent the results from simulations of paper
    I. Different colors 
    indicate different initial $\xg$ as shown in the legend.}  
  \label{fig_bfrt}
\end{figure}

Fig. \ref{fig_bfrt} shows the time evolution of the global ratio of
$f_{\rm b,FG}/f_{\rm b,SG}$.
The global ratio of $f_{\rm b,FG}/f_{\rm b,SG}$ increases with time
due to the combined effect of the more rapid decrease in the number of SG binaries
and FG single stars. As was to be expected, this effect is stronger for
moderately hard binaries but is significant also for the very
compact binaries we have studied in this paper.

As shown in Fig. \ref{fig_bfrr400}, $f_{\rm
  b,FG}/f_{\rm b,SG}$ depends on the distance from the centre and
observations, in general, might not cover the entire cluster.
The bottom panels of Fig. \ref{fig_bfrt} show, for example, the evolution of
$\fbfg/\fbsg$  at distances $0.5R_{\rm h}<r<2.5R_{\rm h}$ where $R_{\rm h}$ is the projected half-mass radius (the radial range is similar to that spanned by data in the observational study of binaries in multiple-population clusters by Lucatello et al. 2015). 
Although the simulations presented here are still
simplified and not aimed at a detailed comparison with observations, it
is interesting to note that the values of $\fbfg/\fbsg$ after a few
initial relaxation times of evolution are in many of our simulations
in general good
agreement with the observational values ($\fbfg/\fbsg
\simeq 4-5$) found by Lucatello et al. (2015) in a similar radial range. The data of Lucatello et al. (2015) did not allow a determination of the binary semi-major axis or mass ratios; assuming a  maximum semi-major axis of 2.3 AU (corresponding to a period of $1000d$), as done in Lucatello et al. to derive one estimate of the binary fraction, would correspond to binaries with a hardness parameter $x$ larger than about 10 (assuming a reference velocity dispersion of 5 km/s, a mass of 0.8 $m_{\odot}$ for each of the binary components and a reference mass for the single stars interacting with the binary equal to 0.5 $m_{\odot}$)
\section{Conclusions}

In this paper, we have studied the evolution of compact binaries in
multiple-population globular clusters. 

A number of theoretical studies of the formation of
multiple-population clusters have predicted that SG stars form in a
compact subsystem in the central regions of a more diffuse FG
cluster. In this paper we have explored the implications of the initial differences
between the SG and FG  structural properties for the evolution of
their compact binary populations.

The binary-binary and binary-single interactions leading to the disruption or ejection
of binaries occur mainly in the cluster central regions where SG binaries are
initially the dominant population and therefore those preferentially affected by
these dynamical processes.

Preferential disruption and ejection of SG binaries are only in part
offset by the preferential relaxation-driven evaporation of FG
binaries and the net result is a more rapid decline in the number of
SG binaries. This effect is qualitatively similar to that found in paper I for
moderately hard binaries, but, quantitatively, is smaller
for the compact binaries considered in this paper.  

For compact binaries, binary-single interactions result in the ejection of both the binary and the single star involved in the interaction; binary ionization and evolution of the binding energy of binaries retained in the cluster are due primarily to binary-binary interactions. We followed the evolution of the binary binding energy distribution and shown that the evolution of the binding energy of interacting binaries is consistent with the predictions of previous binary-binary encounter experiments (Mikkola 1983). 

The evolution of the radial variation of the binary fraction is driven by the interplay between  binary segregation,
ionization and ejection. Segregation leads to an increase of the
binary fraction in the cluster inner regions; ionization and ejection
counteract in part
the effects of segregation but for compact binaries the effects of segregation dominate
and the inner binary fraction increases during the cluster evolution. 
This is different from what happens for the wider binaries considered
in paper I; in that case the larger ionization rate offset the effects
of segregation and resulted in a binary fraction smaller than the initial one
at almost any distance from the cluster centre.

 The local SG-to-FG binary number ratio at a given radius decreases with the distance from the
 cluster centre and is in general different from its global value.
For the compact binaries studied in this paper, the local SG-to-FG binary number ratio is
approximately equal to the global value at a distance from the cluster
centre approximately equal to $0.5r_{\rm h}$.   

Finally we have explored the evolution of the  binary fraction of each population separately (defined as the ratio of the number of binaries in a given population to the total number of binary and single stars belonging to the same population). As a result of the preferential ionization and ejection of SG binaries and the preferential evaporation of FG single stars  the fraction of binaries in the FG population is larger than that of the SG population. The extent of the difference between the FG- and the SG-population binary fraction increases during the cluster evolution and, at a given time, is small in the central regions and increases with the distance from the cluster centre.

A recent study by Lucatello et al. (2015) has shown that
in cluster intermediate regions, the incidence of binaries in the FG population is indeed larger than that of the SG population in general agreement with the findings of this paper and those of simulations presented in paper I.

\section*{Acknowledgements}
EV, JH, and SLWM  acknowledge support by grants NASA-NNX13AF45G and HST-12830.01-A.
This research was supported in part by Lilly Endowment, Inc., through its support for the Indiana University Pervasive Technology Institute, and in part by the Indiana METACyt Initiative. The Indiana METACyt Initiative at IU is also supported in part by Lilly Endowment, Inc.

\end{document}